%% file: main.tex
\documentclass[conference]{IEEEtran}
\IEEEoverridecommandlockouts

\usepackage{cite}
\usepackage{amsmath,amssymb,amsfonts}
\usepackage{algorithmic}
\usepackage{graphicx}
\usepackage{textcomp}
\usepackage{comment}
\usepackage[hyphens]{url}
\usepackage{fancyhdr}
\usepackage{pgfplots}
\pgfplotsset{compat=1.18}
\usepackage{threeparttable, booktabs}
\usepackage{multirow}
\usepackage{xspace}
\usepackage{xcolor}
\usepackage{pifont}
\usepackage{hyperref}
\definecolor{codegreen}{rgb}{0,0.6,0}
\definecolor{codegray}{rgb}{0.5,0.5,0.5}
\definecolor{codepurple}{rgb}{0.58,0,0.82}
\definecolor{backcolour}{rgb}{0.95,0.95,0.92}

\newcommand{\hlsml}{\textsc{hls4ml} \xspace}
\usepackage{subfig}
\usepackage{caption}

\usepackage{tikz}
\tikzset{%
    baseline,
    inner sep=2pt,
    minimum height=10pt,
    rounded corners=5pt  
}
\newcommand{\code}[1]{\mbox{
    \footnotesize\ttfamily
    \tikz \node[anchor=base,fill=black!12]{#1};
}}

\usepackage{xstring}

\begin{document}

\newcommand{\codename}{\textsc{FireBridge}\xspace}
\title{\codename: Cycle-Accurate Hardware + Firmware Co-Verification for Modern Accelerators}

\author{
\IEEEauthorblockN{G Abarajithan, Zhenghua Ma, Francesco Restuccia, Ryan Kastner}
\IEEEauthorblockA{
UC San Diego \\
\{agnaneswaran, zhm007, frestuccia, kastner\}@ucsd.edu
}
}

\maketitle

\newcommand{\speedup}{50$\times$\xspace}

\input{0_abstract}
\input{1_intro}

\input{1.5_background}
\input{2_prior_work}

\input{3_infrastructure}
\input{4_evaluation}

\input{5_conclusion}

\newpage

\bibliographystyle{IEEEtran}
\bibliography{ref}

\end{document}

%% file: 0_abstract.tex
\begin{abstract}

Hardware-firmware integration is becoming a productivity bottleneck due to the increasing complexity of accelerators, characterized by intricate memory hierarchies and firmware-intensive execution. 
While numerous verification techniques focus on early-stage, approximate modeling of such systems to speed up initial development, developers still rely heavily on FPGA emulation for integrating firmware with RTL/HLS hardware, resulting in significant delays in debug iterations and time to market.
We present a fast, cycle-accurate co-verification framework that bridges production firmware and RTL/gate-level hardware. 
\codename enables firmware debugging, profiling, and verification in seconds using standard simulators such as VCS, Vivado Xsim, or Xcelium, by compiling the firmware for x86 and bridging it with simulated subsystems through randomized memory bridges. 
Our approach provides off-chip data movement profiling, memory congestion emulation, and register-level protocol testing, which are critical for modern accelerator verification. 
We demonstrate a speedup of up to \speedup in debug iteration over the conventional FPGA-based flow for system integration between RTL/HLS and production firmware on various types of accelerators, such as systolic arrays and CGRAs while ensuring functional equivalence. 
\codename accelerates system integration by supporting robust co-verification of hardware and firmware, and promotes a structured, parallel development workflow tailored for teams building heterogeneous computing platforms.


\end{abstract}


%% file: 1_intro.tex
\begin{figure}
    \centering
    \includegraphics[width=\linewidth]{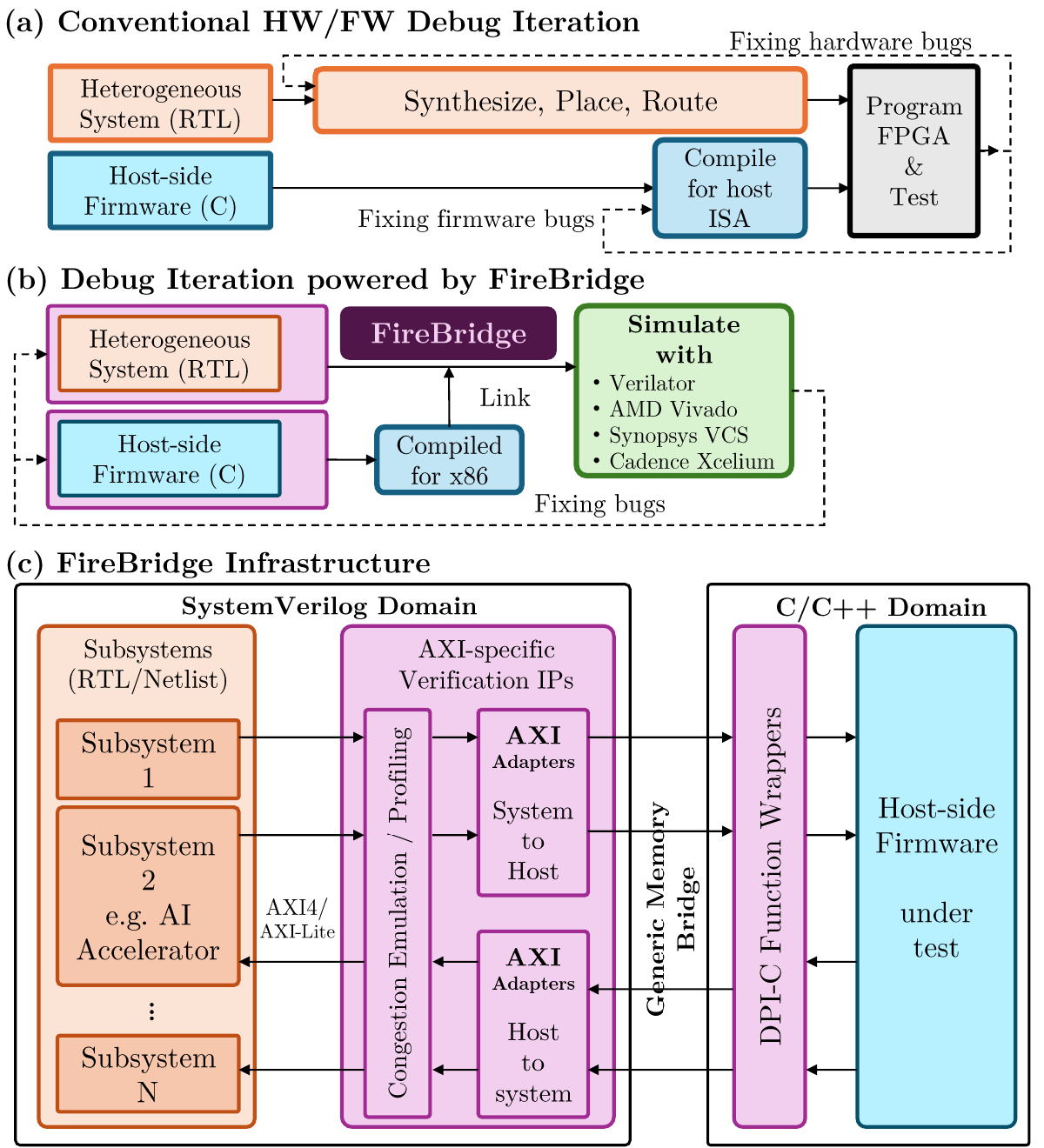}
    \vspace{-5pt}
    \caption{
    \textbf{(a):} Conventional HW/FW integration and debugging flow results in lengthy development cycles. 
    \textbf{(b):} \codename allows the user to compile the firmware natively for x86 and link to the RTL or Netlist to perform behavioral or gate-level simulation through open-source and commercial tools. 
    \textbf{(c):} Our approach wraps host-side firmware in DPI-C interfaces to enable integration with our generalized memory bridges, which interface with transactional SystemVerilog testbenches via an optional memory congestion emulator. }
    \label{fig:infra}
\vspace{-10pt}
\end{figure}

Repository: \href{https://github.com/abarajithan11/axis-systolic-array/tree/master/firebridge}{github.com/abarajithan11/axis-systolic-array}

\section{Introduction}

Rapid advancement of artificial intelligence (AI) has led to the emergence of a wide range of AI accelerators~\cite{dnnacclsurvey}, each tailored for specific applications in domains such as computer vision, natural language processing, autonomous systems, and scientific computing. 
These accelerators, often deployed as heterogeneous IP blocks within System-on-Chip (SoC) platforms, implement highly complex memory access patterns due to the data-intensive nature of deep neural networks (DNNs), which account for over 60–75\% of total latency in practical workloads~\cite{smaug}.
They also pose significant challenges in verification, especially in early-stage firmware development, where understanding full-system interactions is crucial.

The conventional workflow typically involves building the RTL/HLS design of the accelerator, integrating it within an SoC, and then writing firmware and testing it on an FPGA through hardware emulation. 
This process is slow, hardware-dependent, and ill-suited for iterative development. 
Functional bugs in memory protocols or firmware-induced corner cases often only surface after full integration, significantly increasing debug and development cycles. 
Additionally, hardware emulation on the FPGA does not verify the physical design process, such as clock domain crossings, which are often verified independently in gate-level simulations.

To accelerate the development time, prior works, such as Gem5-accel~\cite{gem5acc} and SMAUG~\cite{smaug} have attempted to evaluate simpler, pre-RTL models of accelerators linked to system simulators such as Gem5~\cite{binkert2011gem5}. 
While this is suitable for early-stage design, it does not facilitate the identification of bugs in RTL/HLS implementations of accelerators. 
Utilizing them to verify hardware-firmware integration requires building and maintaining a C++ behavioral model of the accelerator that must stay in sync with RTL throughout the development cycle, and continually ensuring their equivalence doubles the overall verification effort.

Therefore, we present \codename (\textbf{F}irmware-\textbf{i}n-\textbf{R}TL \textbf{E}valuation \textbf{Bridge}), a fast and scalable co-verification framework that directly integrates firmware with the RTL (or gate-level netlist) using SystemVerilog DPI-C. 
The firmware is compiled natively for x86 and is connected to standard memory interfaces, such as AXI, through DPI-C wrappers.
Our method allows production firmware to be tested against real RTL memory accesses and register protocols within seconds, using commercial simulators such as VCS, Xcelium, or Vivado.

We also provide functionality to profile the transactions between the hardware subsystems, DDR memory, and memory-mapped registers.
Off-chip data movement costs orders of magnitude more energy than computation itself~\cite{energyproblem}, and hence accelerators aim to minimize it through extreme data reuse.
The data movement in a real system-on-chip significantly varies from design estimates, due to issues such as stalls and denial-of-service within hierarchical interconnects~\cite{axirealm}~\cite{mutlu2015main}.
Our profiling features help to identify and fix such unexpected issues before deployment.

Our results demonstrate up to \speedup faster hardware-firmware co-verification cycles than FPGA emulation, empowering hardware and software teams to collaboratively iterate and debug new SoC designs with significantly higher productivity.
Conventionally, firmware integration begins after extensive hardware-only verification, which can take weeks to iteratively debug interactions on FPGA-based emulation platforms.
\codename enables such teams to begin firmware development in parallel with RTL/HLS design, iteratively verifying their interactions and fixing bugs in simulation using standard high-performance simulators. 
They would then deploy the same firmware on the emulation platform and get it working within the first few attempts. 
Our key contributions include:

\begin{itemize}
    \item A framework to co-verify, profile, and iteratively develop firmware and heterogeneous hardware (RTL, Netlist) designs, enabling up to \speedup faster debug iterations.
    \item A novel approach of compiling firmware to x86 and bridging with hardware subsystems via DPI-C to accelerate verification time.
    \item Abstractions that ensure firmware and hardware debugged using \codename seamlessly transition to FPGA emulation, significantly reducing integration issues..
    \item Support for profiling off-chip data movement, bandwidth utilization, and accesses to sensitive memory regions.
    \item Memory congestion emulation to stress-test the design and to uncover SoC protocol handling related bugs.
    \item Native compatibility with industry-standard verification tools such as VCS, Xcelium, and Vivado Xsim.
\end{itemize}

%% file: 1.5_background.tex
\begin{figure*}
    \centering
    \includegraphics[width=0.9\linewidth]{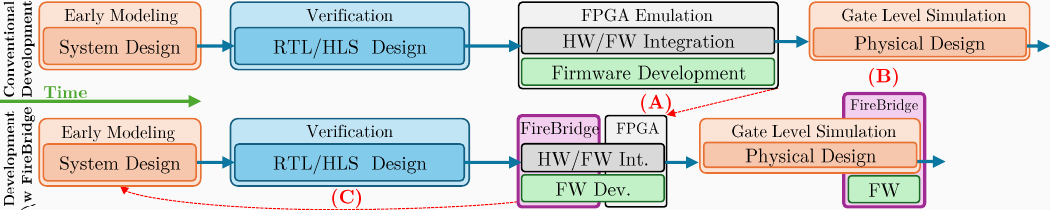}
    \caption{Comparing conventional development of accelerator-based embedded systems and development with \codename. Prior work focuses on early modeling of accelerators. \textcolor{red}{(A)} \codename significantly shortens time spent on firmware development and integration testing with hardware on FPGA by allowing teams to directly test the final firmware in both cycle-accurate RTL/HLS verification. \textcolor{red}{(B)} It also enables testing the final firmware in gate-level simulation. \textcolor{red}{(C)} Our profiling features also allow measuring and debugging data movement issues against those expected from early modeling.}
    \label{fig:workflow}
\end{figure*}

\section{Background}

This section details the steps in the development of an embedded system with an accelerator, unique challenges faced in verifying modern accelerators, a concrete use case of \codename from the perspective of a widely used hardware design framework, and DPI-C, the key ingredient used by \codename to bridge hardware and firmware within verification, while being compatible with all major verification tools.

\subsection{Embedded Accelerator Design Flow}

The four main stages in the development of accelerator-based systems, shown in Fig.~\ref{fig:workflow} are as follows.

\subsubsection{System Design}
System architects utilize frameworks such as Gem5~\cite{binkert2011gem5} or Electronic System Level (ESL) modeling approaches to analyze the feasibility of potential architectures and their dataflow patterns and estimate key metrics, including latency, throughput, and memory bandwidth utilization.

\subsubsection{Hardware Design}
The chosen hardware is implemented as Register Transfer Level (RTL) or using High-Level Synthesis (HLS) tools. 
Its submodules are first tested in smoke tests, then in unit-level randomized testbenches, and finally integrated and tested with system-level testbenches using frameworks such as Universal Verification Methodology (UVM).

\subsubsection{Firmware Development}
The hardware team generates address maps for control \& data registers throughout the heterogeneous design hierarchy and hands them to the firmware team.
Parts of the firmware can be developed in parallel to hardware, but functionality that involves hardware-firmware interactions can only be done after the design is ready for hardware emulation.
Even then, debugging across the hardware implemented in the FPGA and the firmware, as well as finding the root of bugs, is a time-consuming process.

\subsubsection{Physical Design}
For ASICs, the design is then synthesized, placed, and routed for a given technology node, and verified with gate-level simulation before tape-out. 
For FPGAs, the design is further validated before deployment.

\subsection{Complex Memory Access Patterns in Modern Accelerators}

Such hardware-firmware integration has become increasingly challenging as deep learning accelerators adopt more intricate memory access patterns to optimize data movement across their complex memory hierarchies.
The weights, inputs, and outputs of each layer in deep neural networks are multidimensional tensors.
The number of multipliers and accumulators in an accelerator is orders of magnitude less than the number of operations required to compute a layer.
This necessitates unique \emph{dataflow patterns} for each accelerator, where multidimensional tensors are sliced into even more dimensions (tiling), and the order of dimensions is shuffled (N-D transpose) and read from or written to memory out-of-order.
Such complex patterns are adopted to ensure all processing elements are fed in every clock cycle, thereby maximizing the performance efficiency of the accelerator, and to reduce energy consumption by minimizing off-chip data movement through the reuse of data in multiple levels of caches.

A typical accelerator has multiple high-bandwidth memory access ports (e.g. AXI manager ports). 
When integrated into an embedded system-on-chip, they share the data bus with other subsystems (AXI managers and subordinates) in the chip through a hierarchy of memory interconnects.
This makes the data movement non-deterministic, and debugging a complex dataflow on top of that becomes a herculean task.

\subsection{Firmware}

In modern accelerator-based systems, complex data transformations required to facilitate the correct operation of such memory access patterns are handled by the software stack, also known as the firmware \cite{smaug}.
Since accelerators only support specific dataflows, multidimensional input tensors must first be tiled, rearranged, and flattened, where noncontiguous slices of the tensor are copied into contiguous data that can be fed to the accelerator.
Similarly, the output data of the accelerator comes tiled, which should be \emph{untiled} and \emph{retiled} for the next operation or to be given as the output of the inference.
These operations in firmware often account for over 70\% of the inference latency\cite{deepatfb}\cite{smaug}.
\codename provides the profiling tools to measure these bottlenecks, compare against the early-modeling estimates, and fix them before tapeout or deployment.

\input{table_works}

\subsection{FPGA Emulation}

The most common method of system-level integration testing, especially with firmware, is FPGA emulation.
If the goal is to verify the hardware design of the accelerator IP block, an FPGA with a hard processor in it, such as a ZYNQ programmable SoC, could be used. 
The accelerator IP would be connected to the processor via an AXI interconnect, and the corresponding C firmware, which runs on the integrated processor, is developed and tested through numerous iterations on the FPGA board.
Instead, if the goal is to verify a full system-on-chip, the entire system, with the processor, is built as an RTL design, implemented in multiple large FPGAs with proprietary software interconnecting them, and run with the firmware developed and executed within the hardware-emulated processor. 
For large systems, such hardware emulation is the only viable method for conducting end-to-end system testing.

However, a non-trivial bug in an accelerator design that expects data to arrive in a particular order within a fixed latency would likely go unnoticed in the steps of architecture modeling, smoke testing, unit-level randomized verification, and complete accelerator verification.
Such a bug would rarely show up in FPGA emulation, and if it did, it would not be easily reproducible.
Tracing down the source of such a bug often takes weeks, unexpectedly delaying the development cycle \& time-to-market. 
\codename allows the accelerator to be tested with production-ready firmware, with randomized, transactional testbenches, before the FPGA emulation step.
The complex memory access patterns can be verified more easily, and any bugs can be traced easily, either through waveforms or breakpoints in simulation, boosting the development of novel accelerators.

\subsection{HLS4ML}
\label{sec:hls4ml-intro}

As a concrete use case, consider \hlsml\cite{hls4ml2018}, an open-source framework widely used within the scientific computing community for accelerating machine learning models on FPGAs and ASICs. 
Its standard workflow involves generating an HLS-based IP with AXI-Stream interfaces, validating it through C-simulation and subsequently integrating it into an SoC. 
Engineers then construct the SoC by incorporating AXI DMA controllers, additional IP blocks for pre- and post-processing data, and iteratively developing data-movement firmware while testing on FPGAs. 
However, the \hlsml community has identified the absence of a fast, full-system verification methodology encompassing all IPs and the inability to iteratively develop firmware in the loop as significant limitations in their workflows\cite{esp4ml}.
We demonstrate \codename's ability to comprehensively verify \hlsml designs with firmware.

\subsection{Direct Programmable Interface for C (DPI-C)}

SystemVerilog DPI-C (Direct Programming Interface for C) is a powerful interface that allows seamless integration between SystemVerilog testbenches or RTL code and C/C++ functions. 
It enables users to call C functions directly from SystemVerilog, and vice versa, without the need for complex inter-process communication mechanisms. 
This interface is particularly useful for tasks that benefit from C/C++ performance or existing software libraries, such as algorithmic modeling, high-performance data manipulation, or interfacing with system-level software stacks. 
It is much easier to write golden models in C/C++ using existing libraries.
For example, to verify an accelerator for signal processing tasks, one would use FFT libraries, model the accelerator, then import that as a C function into SystemVerilog via DPI-C and compare its results with those of the design under test.
Our key novelty lies in leveraging this feature to address the significant challenge of long debug iterations in accelerator development.

%% file: table_works.tex

\newcommand{\ratingbar}[1]{%
  \begin{tikzpicture}[scale=0.4, baseline=-0.5ex]
    \foreach \i in {1,2,3,4} {
      \pgfmathsetmacro\x{(\i - 1) * 1.5}
      \IfSubStr{,#1,}{,\i,}{%
        \filldraw[fill=black, draw=black] (\x,0) circle (0.5);
        \node[text=white, font=\scriptsize] at (\x,0) {\i};
      }{%
        \draw[fill=white, draw=black] (\x,0) circle (0.5);
        \node[text=black, font=\scriptsize] at (\x,0) {\i};
      }
    }
  \end{tikzpicture}%
}
\newcommand{\circlelabel}[1]{%
  \tikz[baseline=-0.5ex] \node[draw, circle, fill=black, inner sep=0pt, minimum size=1.2em, font=\scriptsize, text=white] {#1};%
}

\begin{table*}[ht]
\centering
\newcommand{\chk}{\checkmark}
\begin{tabular}{@{}lcccccc@{}}
\toprule
                        & \textbf{Simulation Type} & \textbf{Cycle Acc.} & \textbf{Firmware}   & \textbf{RTL Sim} & \textbf{Memory Profiling} & \textbf{Verification Stage\textsuperscript{\textdagger}} \\
\midrule
\textbf{\codename}      & RTL + C FW               & \chk                & Production          & \chk             & Bus level (AXI..)          & \ratingbar{3,4} \\
SMAUG\cite{smaug}       & C++ HW Model + gem5      & -                   & Functional          & -                & Analytical                 & \ratingbar{1}   \\
PARADE\cite{parade}     & Abstract FW + HW         & -                   & Abstracted          & -                & Abstract                   & \ratingbar{1}   \\
STONNE\cite{stonne}     & Cycle-level              & \chk                & -                   & -                & DRAM Trace                 & \ratingbar{1}   \\
SCALE-Sim\cite{scalesim}& Analytical               & -                   & -                   & -                & DRAM Trace                 & \ratingbar{1}   \\
LAMBDA\cite{lambda}     & Analytical               & -                   & -                   & -                & Access counts              & \ratingbar{1}   \\
FireSim\cite{firesim}   & FPGA-Emulated            & \chk                & Production          & \chk             & Full System                & \ratingbar{3}   \\
\bottomrule

\multicolumn{7}{p{0.9\textwidth}}{%
  \footnotesize
  \parbox{\textwidth}{%
    \textsuperscript{\textdagger} Circles correspond to the four stages in the development flow of accelerator-based embedded systems shown in Fig. \ref{fig:workflow}:\\
    \circlelabel{1} Early System Design, \circlelabel{2} RTL/HLS Hardware Design, \circlelabel{3} Firmware Development \& Integration with Hardware, \circlelabel{4} Physical Design.
  }%
}
\end{tabular}
\caption{Comparison of Simulation Frameworks Across Firmware and Hardware Co-simulation Capabilities}
\label{tab:related}
\end{table*}

%% file: 2_prior_work.tex
\section{Related Work}

With increasing interest in deep learning accelerator design, several works have identified full system validation and verification as critical bottlenecks and proposed solutions.
They mainly focus on early-stage, high-level modeling, which involves identifying and resolving as many problems as possible in the first stage (System Design in Fig. \ref{fig:workflow}) of the development process, helping architecture researchers in design space exploration (DSE). 

Table \ref{tab:related} compares relevant related works with our work. 
It can be seen that \codename does not compete with these frameworks, but supplements them, by validating their pre-RTL performance analysis on post-RTL/HLS designs.
A newly designed system still requires hardware-firmware integration via emulation, which some works focus on accelerating, to speed up the integration process.

\subsection{Early-Stage Modeling}
\label{subsec:prior:earlymodel}

Gem5\cite{binkert2011gem5} and its extensions (e.g., gem5-Aladdin, gem5-SALAM, gem5-X) remain popular for early-stage full-system simulation of AI accelerators due to their modularity and support for heterogeneous memory hierarchies. 
Aladdin~\cite{aladdin} provided a trace-based approach to estimate power, performance, and area without requiring RTL, while gem5-Aladdin~\cite{gem5aladdin} extended this capability to simulate complete SoCs by combining Aladdin with gem5. 
SMAUG~\cite{smaug} builds on gem5-Aladdin to offer end-to-end simulation of DNN workloads, incorporating multiple accelerator models and workload schedulers.
The user specifies a DNN with their Python API, and a C++ architecture model, and holistically evaluates the performance of the DNN from the software stack and data movement to the behavior of the hardware accelerator.
This evolution reflects a growing awareness that optimizing just the accelerator microarchitecture in isolation is insufficient.

PARADE \cite{parade} helps system architects by automatically generating accelerator models and running cycle-level simulations with Gem5.
STONNE \cite{stonne} lets the user model an accelerator using a library of micro-architectural elements (adder trees, dataflow type) and simulate them directly from PyTorch.
ASTRA-sim is a cycle‑level\cite{astra}, event‑driven simulator for distributed deep‑learning workloads that integrates accelerator unit models and network models to enable large-model design‑space explorations. 
ScaleSim \cite{scalesim} performs cycle-level simulations based on analytical models of the accelerator and dataflow, without RTL/HLS.
NNASIM\cite{nnasim} is a fast, event-driven simulator designed for vector-based DNN accelerators for RISC-V based systems. 
It uses a library of building blocks characterized with ASIC tools to derive the timing, and simulates it with RISC-V binaries.
LAMDA\cite{lambda} extends Timeloop\cite{timeloop} / Accelergy\cite{accelergy} to add detailed analytical modeling of communication and memory subsystems, and enable multi-chip module (MCM) accelerator simulation.
ASH\cite{ash} accelerates RTL simulation itself by compling RTL into dataflow tasks and using their custom ASH hardware to accelerate them.

However, in such early-stage modeling, firmware is not executed against actual memory-mapped RTL but instead interacts through high-level abstractions, which limits its fidelity for system-level verification.
While these tools offer fast evaluations, they cannot verify real RTL behavior or register-level protocol correctness. 
Instead, they model data movement and resource contention, and assume fixed-function accelerators, requiring detailed workload traces. 

\subsection{SystemC-based Modeling \& Firmware testing}

Electronic System Level (ESL) modeling and verification is another early-modeling paradigm used in the industry to build SoCs.
CoFluent Studio by Intel \cite{cofluent} is a popular industry tool for DSE and modeling. 
HEPSIM2 \cite{hepsim} is an advanced ESL tool that employs SystemC for functional and timing HW/SW co-simulation and analysis.
It utilizes a Communicating Sequential Processes (CSP) model of computation, enabling system behavior representation, simulation, and formal analysis. 
Firmware integration within this ESL approach involves interactions between modeled firmware behaviors and targeted hardware architectures defined during the DSE process. 
Tools such as HEPSYCODE-RT \cite{hepsycodert} and HEPSIM2 enable mapping of software processes onto hardware components, considering real-time and non-functional constraints. 
Such tools do not provide cycle-accurate verification or work on RTL/HLS hardware designs, and cannot be used to identify firmware-hardware integration bugs introduced by subsystems such as SoC interconnects and DMAs.

\subsection{Accelerating FPGA Emulation}

Recent works have also identified emulation as a particular bottleneck and focused on accelerating it.
FireSim\cite{firesim} is an FPGA-accelerated, cycle-exact simulator capable of modeling high performance network \& I/Os with an integrated firmware stack on Amazon cloud FPGAs. 
In contrast, \codename focuses on smaller, accelerator-based embedded systems that can be simulated by the user on their own machines.
Chipyard\cite{amid2020chipyard} harnessed FireSim for full system simulation and validation. 
Farshchi et al\cite{farshchi2019integrating} integrated an NVIDIA Deep Learning Accelerator with RISC-V SoC using FireSim, as an example of incorporating AI/ML accelerators into real-time embedded systems.
RoSÉ\cite{Rose2023} integrates FireSim with AirSim for robotics SoC evaluation, enabling real-time hardware-software interaction through a synchronization bridge.

%% file: 3_infrastructure.tex
\section{\codename}

\codename is built as a randomized transactional verification environment in SystemVerilog (SV) and C, designed to accelerate the functional verification of heterogeneous SoCs. 
As shown in Fig.\ref{fig:infra}(c), the framework consists of SV and C domains, bridged through the DPI-C (Direct Programming Interface for C). 
During RTL or gate-level simulation, the host code is compiled into an x86 binary and linked with the testbench.
This architecture enables efficient execution of host code while maintaining equivalence with the real hardware-firmware system that would run on FPGA/ASIC.

\begin{figure}
    \centering
    \includegraphics[width=1\linewidth]{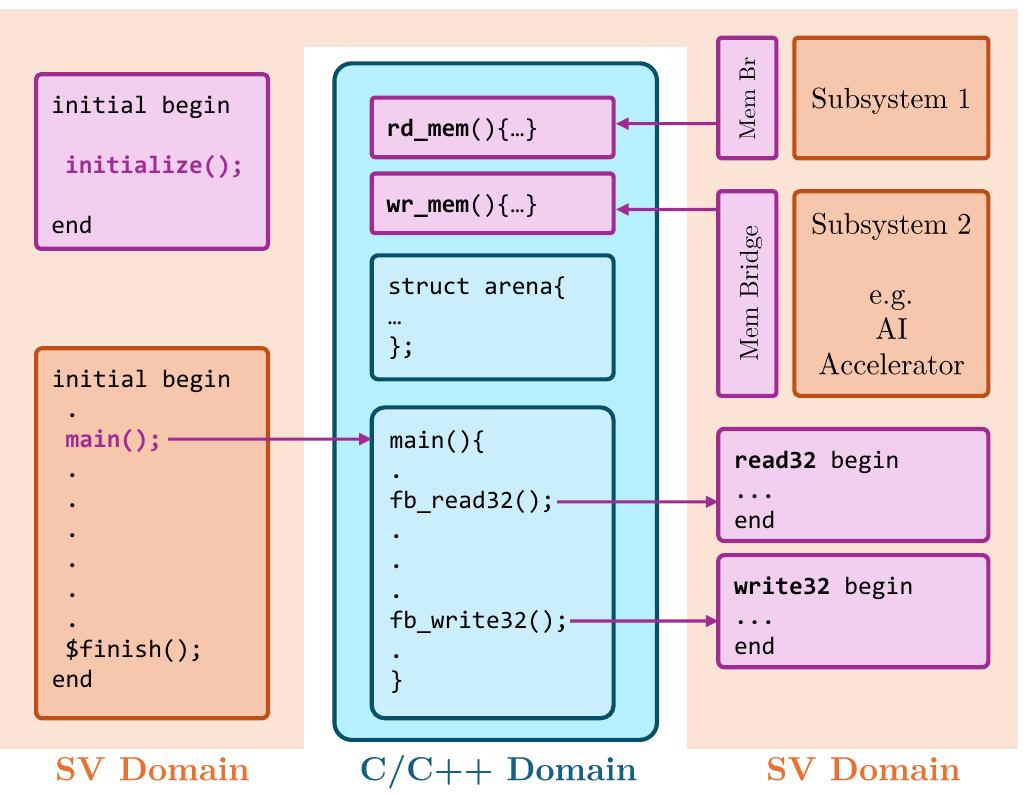}
    \caption{Execution flow of hardware \& firmware bridged via \codename during verification. Control flow begins in a standard SystemVerilog testbench where all initial blocks start together and execute in parallel. \codename is initialized as one of them. The primary initial block starts the firmware.Reading/writing hardware registers is done through the provided functions. DDR can be accessed through regular C pointers. Hardware subsystems, such as accelerators, access the memory maintained in the C domain through provided generic bridges.}
    \label{fig:execution}
\end{figure}

The control flow of the entire simulation is described in Fig \ref{fig:execution}. It begins within the \code{initial begin..end} blocks of the SV testbench, to ensure native compliance with industry tools.
The user wraps their host-side firmware in C functions provided by \codename, which get imported as SV tasks and called within the main \code{initial} block.
The DDR memory of the overall system under test is mapped to the DDR of the user's machine and maintained within the C domain for maximum performance.
A core strength of our framework is its full compatibility with various commercial and open-source simulation tools, such as Verilator, AMD Vivado Simulator, and Cadence Xcelium.

\subsection{User Workflow}

Our API is designed to test the final firmware in the verification stage, minimizing the overhead required to interface with DPI-C, boosting the productivity of embedded engineering teams.
The user's host-side firmware reads and writes to DDR memory using idiomatic C, by dereferencing pointers and accessing arrays.
Registers and on-chip memory that are memory-mapped can be accessed in the firmware through \code{fb\_read\_32(addr)} and \code{fb\_write\_32(addr, data)} functions provided in \codename.
This abstraction enables the firmware to execute control flows seamlessly, such as writing to control registers to trigger hardware actions, polling status registers for synchronization, and post-processing data generated by other subsystems. 
These functions are SV tasks that interface with our generic memory bridges, which in turn interface with subsystems in the design under test via bus-specific (e.g., AXI) VIPs.
This allows \codename to monitor and profile the transactions between the hardware and firmware without sacrificing performance.
After verification, when the same host code is compiled for the target architecture, such as ARM or RISC-V, the wrapper functions of \codename are set to be statically optimized away, producing the same functionality in the taped-out ASIC or the final FPGA-based system-on-chip. 





\subsection{Cycle Accuracy}

Many early-stage modeling tools, such as those based on Gem5 provide cycle-level simulation of the CPUs, but rely on analytical models for the custom hardware.
However, when you are building the custom hardware yourself and actively debugging it, cycle-level accuracy is critical for RTL simulation. 
At the gate level, even greater precision is required to uncover subtle issues, including sub-cycle effects like parasitic delays.
Achieving that level of precision in CPU simulation would be prohibitively time-consuming. 
Moreover, since CPU IPs are typically well-validated and come with high licensing costs, such detailed simulation is generally unnecessary.
As \codename intends to speed up hardware-firmware integration cycles, we choose to maintain cycle (and sub-cycle) accuracy for RTL while maintaining functional equivalence of the firmware.

\subsection{Memory Bridges}

The subsystems on the SV domain can use any bus interface. 
Our memory bridges interfaced with the DPI-C wrappers are protocol-independent, supporting protocols such as AXI and TileLink. 
We provide wrapper verification IPs for AXI4 as an example.
Requests initiated from either side trigger DPI-C calls, enabling seamless memory access synchronized with the hardware simulation.

We include a model within the framework to emulate extreme bus congestion behavior. 
This allows randomized control of memory access signals with adjustable probabilities while adhering to the protocols, allowing the user uncover hardware bugs related to incorrect bus protocol handling. 

\subsection{Profiling \& Monitoring Features}

\codename also optionally profiles bus utilization, memory stalls, and memory access patterns over the simulation time.
This helps users benchmark the final RTL/HLS hardware and final firmware against the numbers predicted by pre-RTL modeling frameworks described in Section \ref{subsec:prior:earlymodel}, and validate performance before moving on to FPGA deployment or ASIC physical design flow.
We demonstrate this feature in Fig. \ref{fig:graph_bandwidth} and Fig. \ref{fig:graph_heatmap}, on a firmware-heavy accelerator running a CNN.

%% file: 4_evaluation.tex
\begin{figure}
    \centering
    \includegraphics[width=\linewidth]{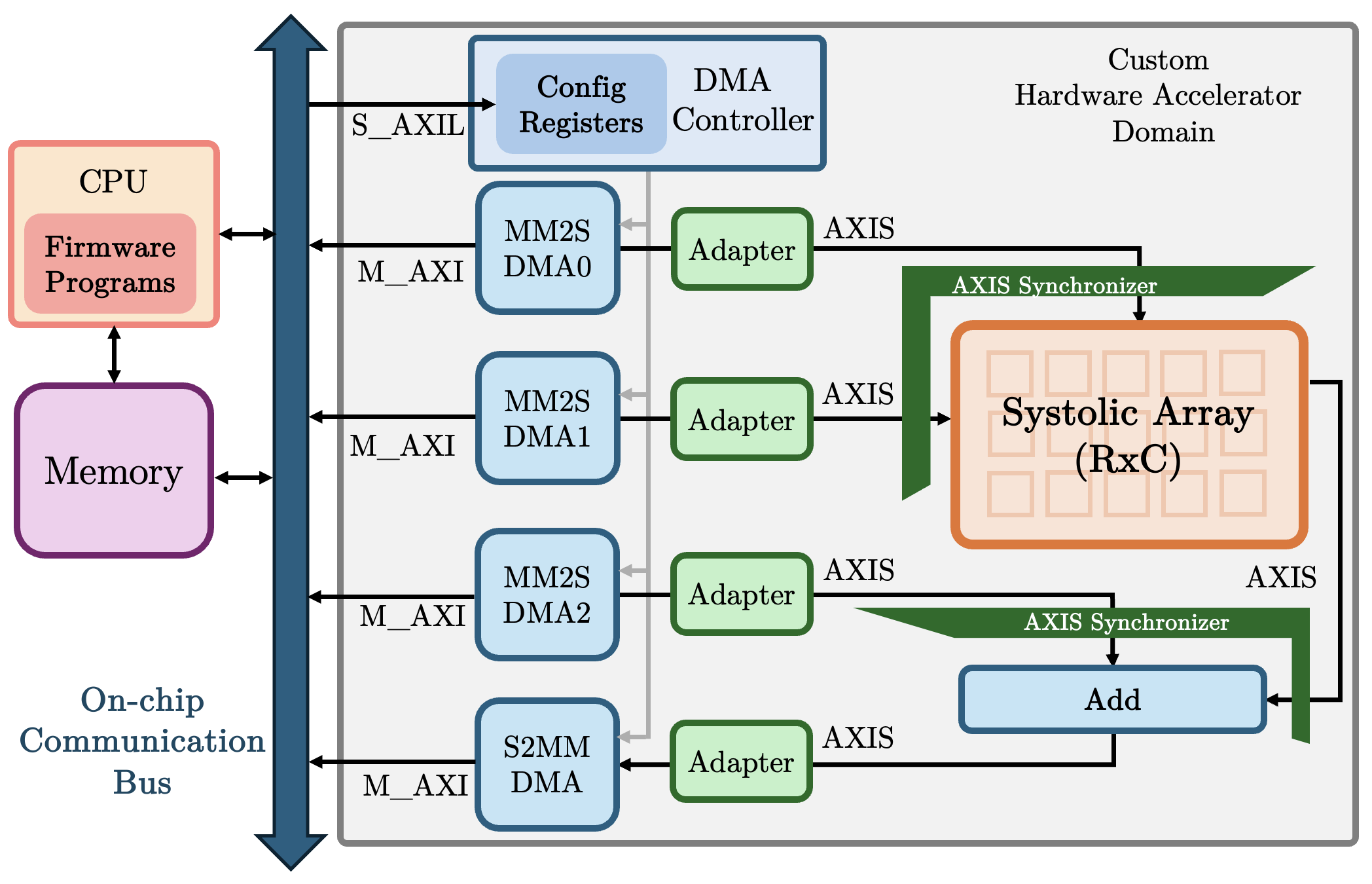}
    \caption{The representative accelerator-based system used to compare the time \& resources required for traditional vs proposed hardware-firmware integration flows. A 2D systolic array of 8-bit multipliers and 32-bit accumulators is integrated with four AXI4 DMAs.}
    \label{fig:systolic}
\end{figure}

\begin{figure}
    \centering
    \includegraphics[width=1\linewidth]{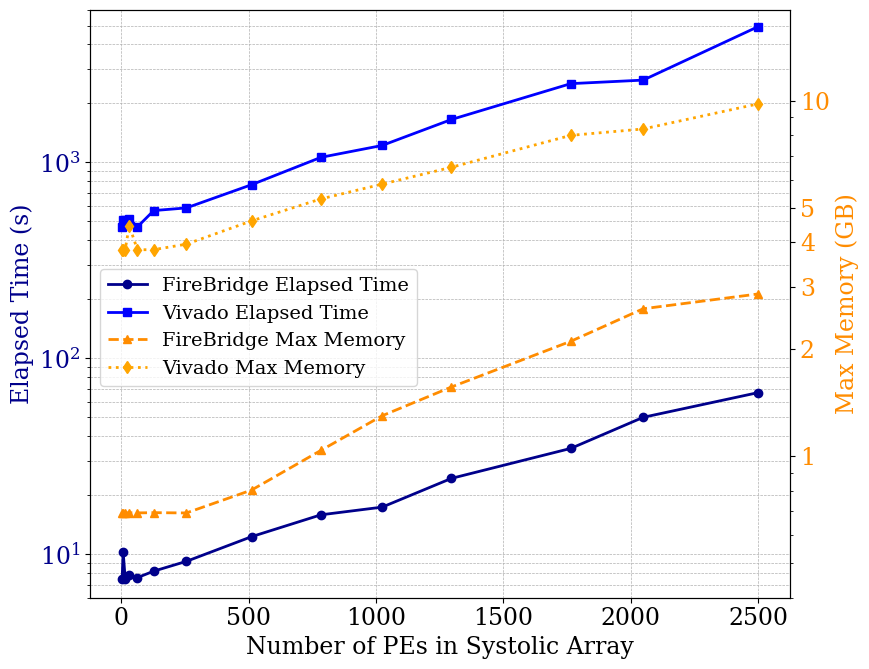}
    \caption{Comparing the runtime of FPGA implementation flow using Vivado, vs simulation flow with \codename, as a proxy for the time user spends on an iteration of debugging hardware-firmware integration. A representative SoC (Fig. \ref{fig:systolic}, Section \ref{subsec:systolic}) with a systolic array of varying numbers of processing elements is taken through both flows to demonstrate the scalability of \codename with increasing design complexity. 2500 PEs is the largest design that fits in the FPGA.}
    \label{fig:graph_systolic}
\end{figure}

\section{Evaluation}

We evaluate both the traditional flow and our proposed flow on various types of accelerators \& corresponding firmware to demonstrate the productivity boost enabled by \codename.
A representative SoC with multiple AXI4 DMAs and a systolic array of varying sizes is taken through both flows to characterize their performance scaling with design complexity, measured in FPGA resources.
SoCs with end-to-end neural networks are built using \hlsml and evaluated in both flows to demonstrate the utility of \codename for HLS-based designs and the scientific computing community, solving a real-world problem.
We utilize AMD ZCU102~\cite{zcu102}, an FPGA evaluation kit featuring 600K logic cells, 2.5K DSPs, and an integrated ARM Cortex-A53 processor, for hardware-firmware integration and debugging of the entire system.


\subsection{Traditional vs Proposed Debug Iterations}

This section provides a brief description of the proposed flow and compares it with a traditional development flow in a typical hardware acceleration scenario for AMD platforms.
We take different kinds of representative accelerators, described in Subsections \ref{subsec:systolic}, \ref{subsec:hls5ml}, and \ref{subsec:cgra4ml} for a fair comparison.


\subsubsection{Traditional Debug Iteration} 
We generate a block diagram \& Xilinx IPs, synthesize, place, and route the design on the FPGA fabric using AMD Vivado 2020.2.
We then use AMD Vitis 2020.2 to build and deploy the firmware and bitstream on the FPGA board.
At this step, users face the firmware-hardware integration bugs that frequently arise from the complex data movement patterns of the accelerators.
For example, the AXI DMAs often \emph{hang} / become unresponsive when the amount of data requested or the timing between DMAs is incorrect, and the memory-mapped registers usually do not read/write data correctly.
To debug such errors, a user typically marks specific signals for debugging, invokes the Internal Logic Analyzer (ILA), synthesizes, places, and routes the design, and deploys firmware \& bitstream to the FPGA again, and observes the behavior.
This is repeated for days, or often weeks, until all bugs are identified and fixed in hardware and firmware. 
For a conservative estimate, we measure the runtime of Vivado to synthesize, place \& route the system with ILAs placed on two AXI buses, as the debug iteration time for FPGA emulation.

\subsubsection{Proposed Debug Iteration}
We integrate the RTL/HLS of the system and the firmware using \codename, and run Vivado Xsim to simulate both together.
The user would encounter integration bugs at this step before proceeding to the FPGA.
To debug such errors, the user would view the VCD waveform of all signals (not the limited ILA signals), identify the bug, update the firmware C code and the hardware RTL/HLS code, and then rerun the simulation.
After repeated debugging iterations, the user would take the firmware and hardware through Vivado + Vitis SDK once, deploy it on the FPGA, and get it working within the first few attempts, as we did. 
This becomes possible due to the equivalence ensured by \codename.
We measure the compile time + runtime of the simulation of RTL/HLS bridged with C firmware via \codename as the debug iteration time for the proposed flow.

\subsection{A Representative System-on-Chip}
\label{subsec:systolic}

Systolic arrays are at the heart of most of the DNN accelerators from the industry, such as Amazon Titanium \cite{amazon_titanium}, Google TPU\cite{google_tpu}, Intel Gaudi \cite{intel_gaudi} \& Tenstorrent Greyskull \cite{tenstorrent}, and from academia, such as AutoSA \cite{autosa} and Gemmini\cite{gemmini}.
Therefore, we demonstrate the scalability of \codename through the verification of data access patterns and firmware on a representative system-on-chip detailed in Fig. \ref{fig:systolic}.
The system consists of a parameterizable systolic array, four AXI4 DMAs, AXI-Stream adapters, and a DMA controller with memory-mapped configuration registers.
The memory-mapped to stream (MM2S) DMAs read data from off-chip DDR through 128-bit high-performance AXI4 ports, and generate full-performance AXI-Stream packets.
The AXI-Stream packets of weights \& input activation are synchronized and sent to the systolic array.
Its output AXI-Stream packet is summed with the partial sum stream and sent to the stream-to-memory-mapped (S2MM) DMA, which converts it into full-performance AXI4 burst transactions that write data into the off-chip memory.
Fig. \ref{fig:graph_systolic} shows the conventional debug iteration and \codename compared on increasingly complex versions of this design until the FPGA is filled.

\subsection{End-to-end DNNs with \hlsml}
\label{subsec:hls5ml}
\begin{figure}
    \centering
    \includegraphics[width=0.8\linewidth]{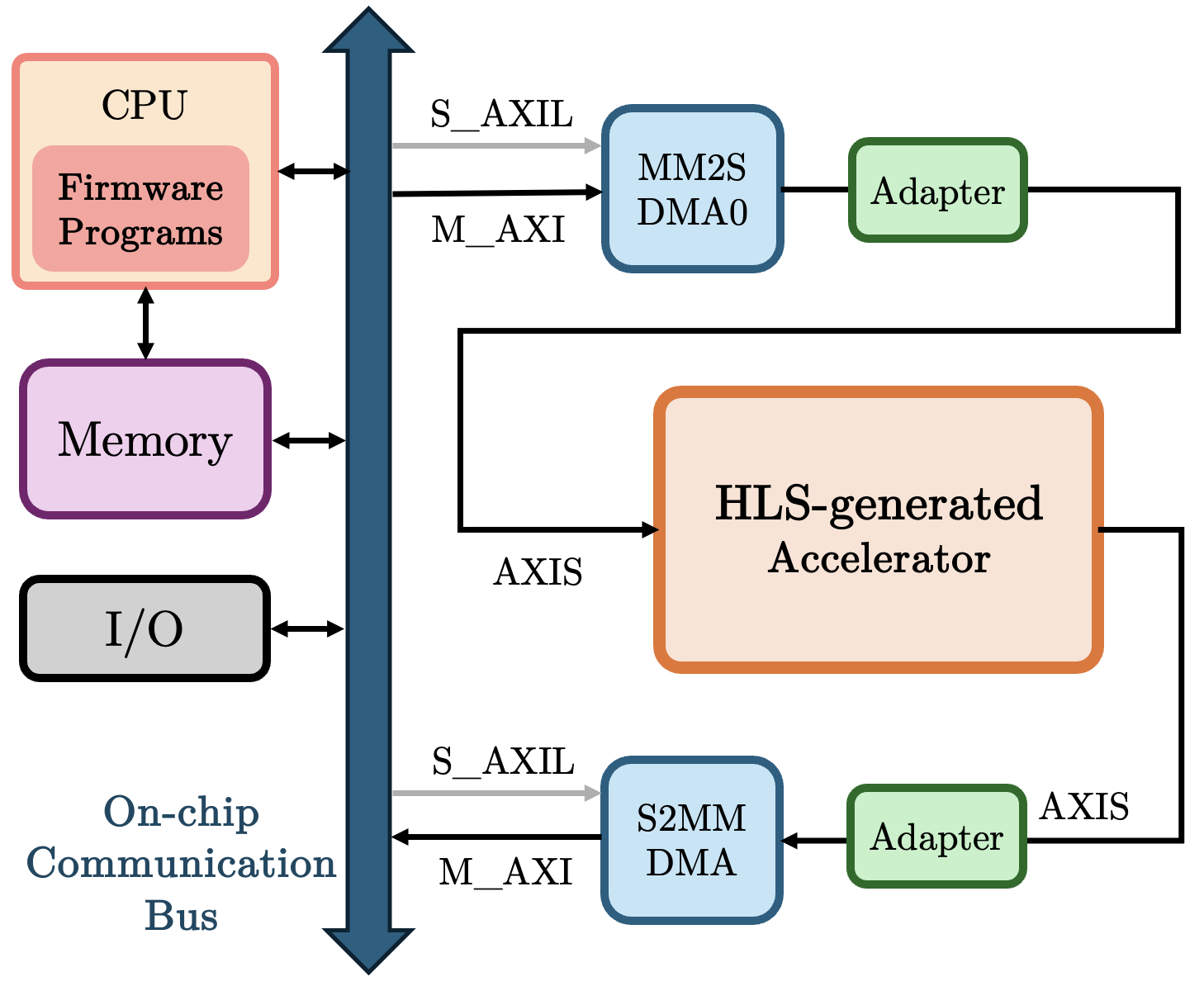}
    \caption{
    An end-to-end, low-latency accelerator system, built using \hlsml~\cite{hls4ml2018} and tested with traditional and proposed flows, to demonstrate the generalizability of \codename, and its utility to the scientific computing community.}
    \label{fig:hls4ml-arch}
\end{figure}

\begin{figure}
    \centering
    \includegraphics[width=\linewidth]{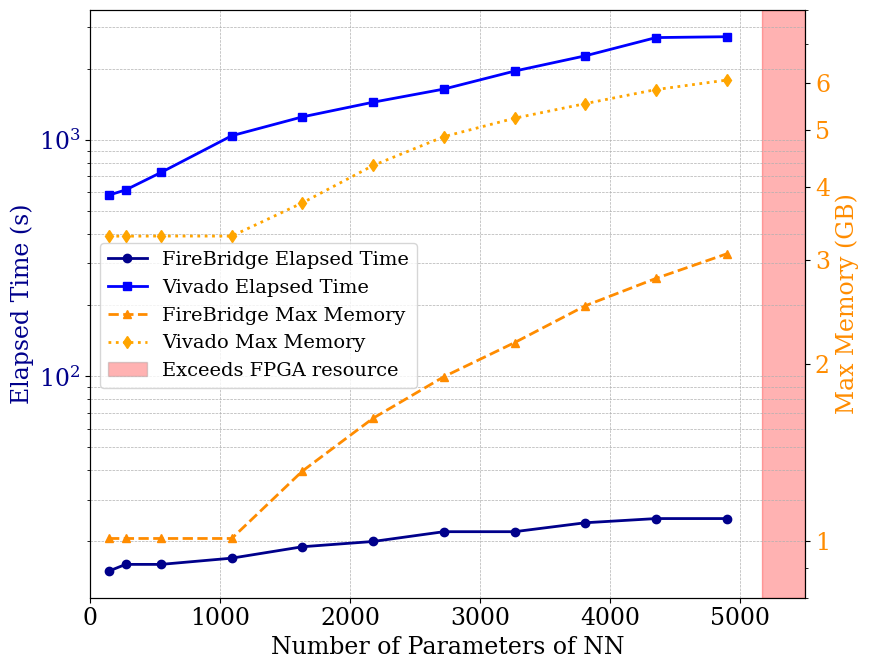}
    \caption{Execution time and max memory usage (log scale) of simulation by \codename vs FPGA prototyping of cascaded dense layers quantized to 16 bits. RTL is auto‑generated by \texttt{hls4ml}. We scale the network until it fails to fit on the ZCU102, which is highlighted in pink, then compare the host‑side runtime and peak memory usage of \codename{} against the FPGA‑prototyping EDA flow.}
    \label{fig:graph_hls4ml}
\end{figure}

We shift focus to a different class of accelerators in this subsection. 
\hlsml, described in  Section \ref{sec:hls4ml-intro}, is an end-to-end toolchain for low-latency DNN accelerator implementation, widely used by the scientific computing community, especially by hardware non-experts. 
Instead of requiring manual RTL design, it automatically generates an HLS-based description of the accelerator from a DNN described in Python. 
The complete SoC architecture produced by HLS4ML is shown in Fig. \ref{fig:hls4ml-arch}, which features a simple memory access pattern with light-weight firmware logic and highly customized, complex hardware logic. 
We conducted experiments using different sizes of DNNs as shown in Fig. \ref{fig:graph_hls4ml}.  The elapsed time and peak memory consumption were measured after generating the RTL using the HLS tools. \codename achieves significantly lower runtime and peak memory usage than the FPGA‑prototyping flow across the hls4ml streaming interface and for all network sizes that still fit on the ZCU102.

\begin{figure*}
    \centering
    \includegraphics[width=1\linewidth]{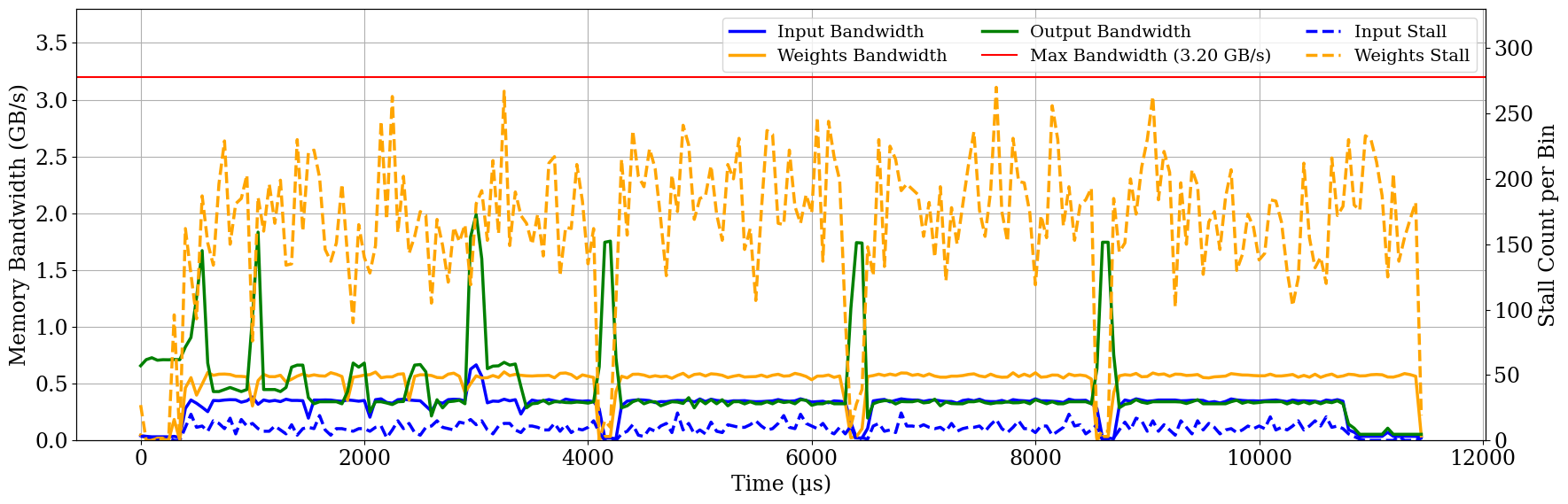}
    \caption{Profiling the memory bandwidth utilization of a CGRA-based accelerator, through the inference of ResNet-18 (0.7B operations). Memory stalls (denial of service by SoC interconnect) are also measured, which helps to identify bottlenecks and fix them before deployment.}
    \label{fig:graph_bandwidth}
\end{figure*}

\begin{figure}
    \centering
    \includegraphics[width=1\linewidth]{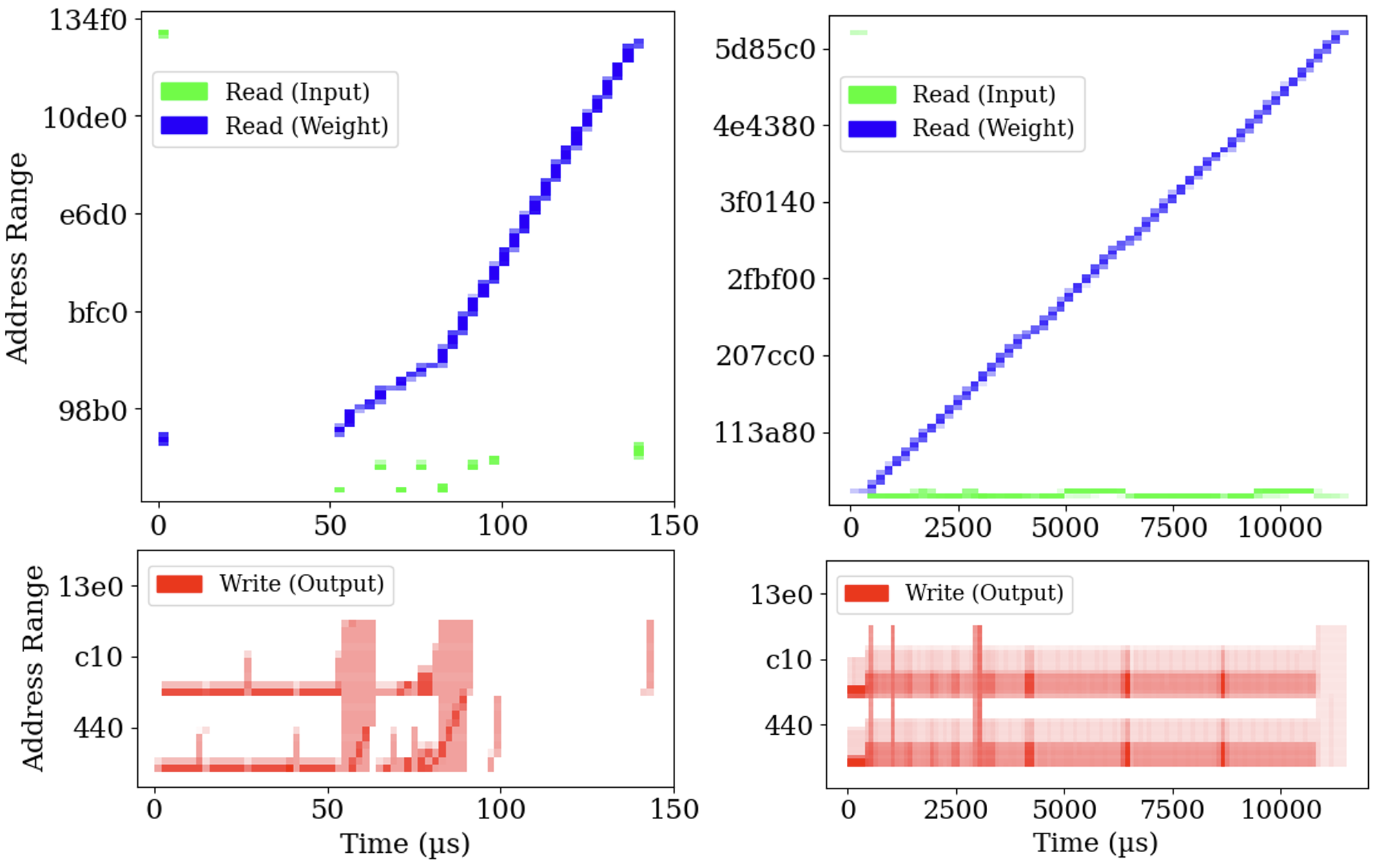}
    \caption{Profiling the memory access patterns of a CGRA-based accelerator, along with an end-to-end inference of a small CNN (left) and ResNet-18 (right) running ResNet-18. }
    \label{fig:graph_heatmap}
\end{figure}

\subsection{End-to-end Evaluation of a Firmware-Heavy Accelerator}
\label{subsec:cgra4ml}
We profile a CGRA-based accelerator using \codename to debug and profile the numerous hardware-firmware interactions required to run the inference of an entire model.
The accelerator performs heavy operations, such as convolution and matrix multiplication in the CGRA, and pointwise operations and data transformations in the C firmware. 
We use 
Fig. \ref{fig:graph_bandwidth} analyzes the memory bandwidth utilization of the CGRA and its three AXI DMAs over an inference of ResNet-18, a CNN with 0.7 billion operations. 
We can see that the memory bandwidth is significantly underutilized, leaving room for improvement in the SoC architecture.
Since the weights are prefetched, the input DMA was given higher priority based on early-stage modeling analysis.
Therefore, the higher number of stalls faced by weights DMA from the interconnect validates the design trade-offs.
Fig. \ref{fig:graph_heatmap} are heatmaps that show the memory access patterns on a small and a large DNN.
Ping pong buffering on alternating layers can be observed in the \emph{input read} operations, and a consistent pattern of \emph{weights reading} can also be seen.
These observations of hardware-firmware interactions offer critical insights to further optimizing the memory hierarchy, such as adding separate caches for the subsystems that clearly operate on distinct memory regions.

%% file: 5_conclusion.tex
\section{Conclusion}

In this work, we introduce \codename, a fast and cycle-accurate framework for co-verifying firmware and heterogeneous RTL/Netlist hardware systems. 
Leveraging SystemVerilog DPI-C and a protocol-compliant memory congestion emulator, \codename enables production firmware to interact with RTL hardware in simulation, thereby minimizing the time spent in traditional FPGA-based emulation. 
This drastically shortens integration debug cycles by up to \speedup and allows firmware teams to test and refine software in tandem with hardware development.

Our evaluation across diverse accelerator types, including systolic arrays, CGRAs, and widely used accelerator generators, demonstrates the versatility, scalability, and profiling capabilities of our framework. 
\codename bridges the critical gap between high-fidelity RTL testing and production firmware execution, bringing software-in-the-loop verification to the forefront of modern SoC design. 
We plan to release this as an open-source framework to encourage adoption, reproducibility, and community-driven improvements.
By enabling fast, reproducible, and transparent functional integration, \codename facilitates a more systematic and concurrent co-development process between firmware and hardware, addressing key challenges in the design of increasingly heterogeneous architectures.
